\documentclass[conference,10pt]{IEEEtran}

\usepackage{amsmath,graphicx,subfigure,epsfig,amssymb,amsmath,epstopdf,tikz,float,tikz-3dplot,color,pdfpages,cite,booktabs,psfrag,multicol}
\usepackage{import}
\usepackage{dsfont}
\usepackage{url}	
\usepackage{algorithmic}
\usepackage{algorithm}
\usepackage{multirow}
\usepackage{graphicx}
\usepackage{geometry}
\usepackage[T1]{fontenc}

\usepackage{fancyhdr}
\fancypagestyle{firstpage}{%
    \fancyhf{} 
    \fancyhead[L]{\small \textcolor{blue}{This is a preprint version. The published version will be available in the proceedings of IEEE NCC 2025 on IEEEXplore subsequently.}}
}

\usepackage{textcomp}
\usetikzlibrary{decorations.pathreplacing,decorations.pathmorphing,decorations.markings,shapes,backgrounds,patterns,decorations.text}

\title{Novel Closed Loop Control Mechanism for Zero Touch Networks using BiLSTM and Q-Learning}

\author{Tamizhelakkiya K$^{1}$, Dibakar Das$^{1}$, Jyotsna Bapat$^{1}$, Debabrata Das$^{1}$, and  Komal Sharma$^{2}$\\
$^{1}$Networking and Communication Research Lab,  IIIT Bangalore, India\\
$^{2}$Toshiba Software (India) Private Limited, Bangalore, India}

\providecommand{\keywords}[1]{\textbf{\textit{Keywords---}} #1}

\begin{document}
 \maketitle
  \thispagestyle{firstpage}

\begin{abstract}

As networks advance toward the Sixth Generation (6G), management of high-speed and ubiquitous connectivity poses major challenges in meeting diverse Service Level Agreements (SLAs). The Zero Touch Network (ZTN) framework has been proposed to automate and optimize network management tasks. It ensures SLAs are met effectively even during dynamic network conditions. Though, ZTN literature proposes closed-loop control, methods for implementing such a mechanism remain largely unexplored. This paper proposes a novel two-stage closed-loop control for ZTN to optimize the network continuously. First, an XGBoosted Bidirectional Long Short Term Memory (BiLSTM) model is trained to predict the network state (in terms of bandwidth). In the second stage, the Q-learning algorithm selects actions based on the predicted network state to optimize Quality of Service (QoS) parameters. By selecting appropriate actions, it serves the applications perpetually within the available resource limits in a closed loop. Considering the scenario of network congestion, with available bandwidth as state and traffic shaping options as an action for mitigation, results show that the proposed closed-loop mechanism can adjust to changing network conditions.
Simulation results show that the proposed mechanism achieves 95\% accuracy in matching the actual network state by selecting the appropriate action based on the predicted state.

\end{abstract}
\section{Introduction}
With the emergence of Sixth Generation (6G) networks, ``connected intelligence” is paramount and shifts from the paradigm of ``connected things” \cite{khan2020}. Automation and intelligence are crucial to optimize the high-performance demands (bandwidth allocation, low latency, high speed) of 6G applications. Additionally, the networks must overcome challenges, such as link failures, congestion, and security attacks \cite{Fu2018}. The concept of Zero Touch Network (ZTN) \cite{Chergui2022} envisages creating an intelligent, automated, self-configuring, self-monitoring, and self-optimizing network architecture that minimizes human intervention. These advancements align with recognized standards, like the European Telecommunications Standards Institute (ETSI) Zero-touch Service Management (ZSM) specifications \cite{ETSI2019}. The ZTN can be implemented using different methods, such as programmatic control loops, virtualization, orchestration, data analytics, closed-loop control, and Artificial Intelligence (AI) / Machine Learning (ML) \cite{Coro2022}. Automation with AI/ML enables decision-making in congestion control, fault detection, and intrusion detection \cite{CERDAALABERN2023}.

Despite the proliferation of AI/ML in networks for automation, developing and deploying ML models still has many open challenges. So, Automated Machine Learning (AutoML) has been introduced in networks which simplifies the creation of ML models for ZTN security \cite{yang2024}. In \cite{Rani2023}, ZTN architecture has been proposed for network slicing to optimize resource management. However, it lacks efficient End-to-End (E2E) network slicing management. The Zero-Load scheme \cite{Bhattacharya2023} has been investigated for load balancing and anomaly detection in ZTN core routers. Most of the existing works \cite{CERDAALABERN2023} do not implement a complete automated closed-loop control needed to continuously optimize network resources.
The motivation for this work is to design a comprehensive closed-loop control mechanism that effectively addresses these challenges. This system design should include continuous monitoring, analysis, decision-making, and evaluation of actions under fluctuating network conditions.

\begin{figure}[htbp!]
    \centering
     \includegraphics[width=2.8in]{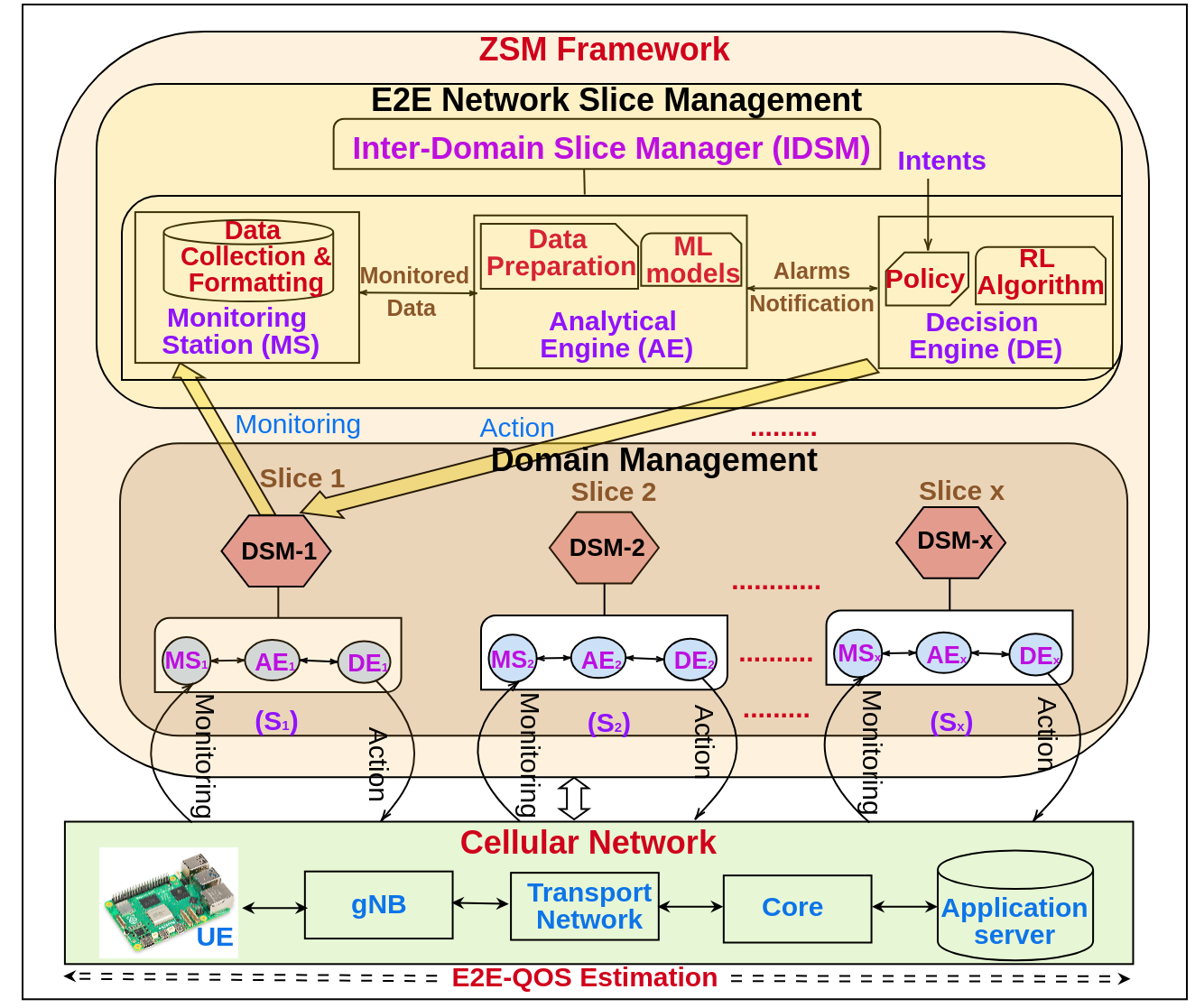}
    \caption{ZTN framework implementation using AI/ML}
    \label{ZTN Framework}
\end{figure}

The ZTN framework \cite{ETSI2019} enables closed-loop operation, as shown in Fig. \ref{ZTN Framework}, and is structured into two management sections: E2E-Network Slice Management (E-NSM) and Domain Management (DM). The Inter-Domain Slice Manager (IDSM) in E-NSM coordinates interactions across network slices, with each slice functioning as an independent virtual network tailored to specific use cases. Each slice includes three components: the Monitoring Station (MS) for real-time performance data collection, the Analytical Engine (AE) for identifying Quality of Service (QoS) issues, and the Decision Engine (DE) for recommending corrective actions. In DM, the Domain Slice Manager (DSM) oversees monitoring and executes the actions recommended by the DE on the cellular infrastructure.

In cellular networks, multiple User Equipment (UE), such as smartphones or Internet of Things (IoT) devices, connecting to various Application (App) servers may experience network congestion.

During network congestion, high traffic loads resulting in reduced performance and significantly increased delays. Internet Service Providers (ISPs) \cite{Rint2022} use various Congestion Avoidance (CA) techniques, such as Traffic Shaping (TS), QoS model selection (e.g., Best Effort (BE), Real-Time Polling Service (RTPS), and Unsolicited Grant Service (UGS)), Content Delivery Networks (CDNs) and load balancing to alleviate congestion and enhance network performance.

\begin{figure}[ht!]
\centering
\includegraphics[width=2.85in]{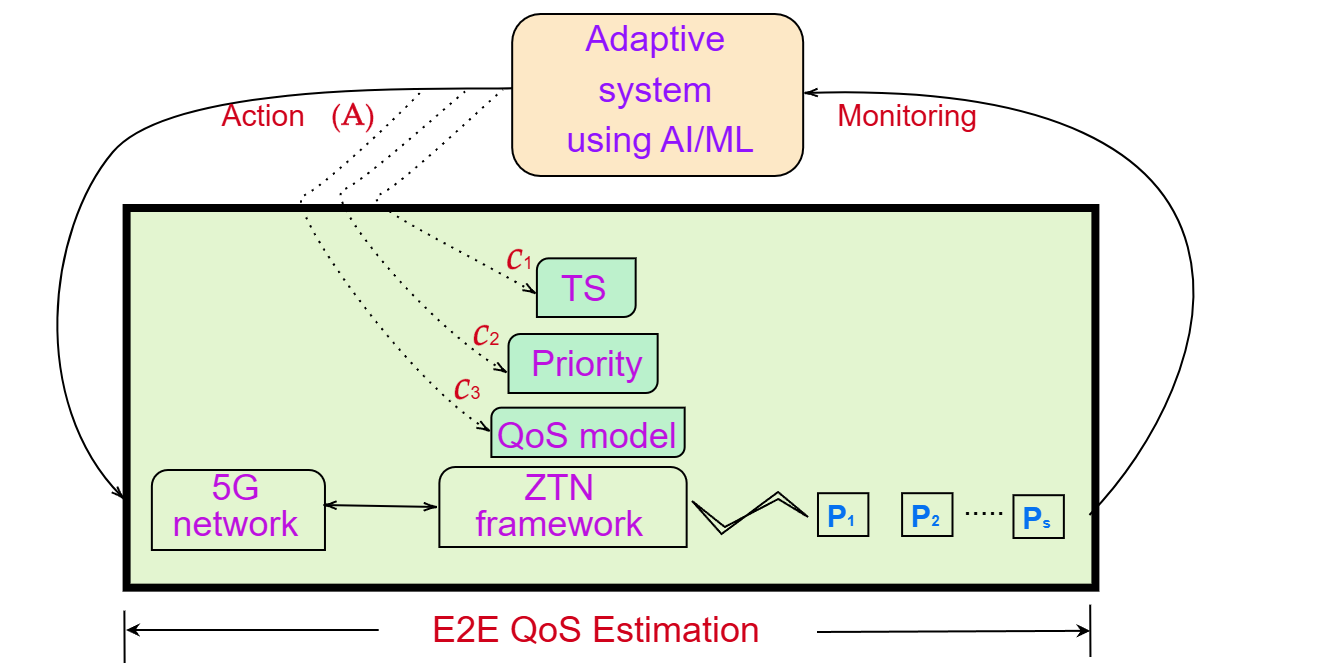}
\caption{Congestion mitigation with closed-loop control}
\label{congestion_control}
\end{figure}

The foundation of efficient network resource utilization during congestion lies in the following key TS parameters:
\begin{itemize}
    \item \textit{Committed Information Rate (CIR)} guarantees a minimum data rate for high-priority services maintaining baseline bandwidth for critical applications under heavy traffic.

    \item \textit{Excess Information Rate (EIR)} allocates data rate to less critical applications by using unused network resources beyond the committed rate.

    \item \textit{Excess Burst Size (EBS)} is the maximum burst of data that can exceed the CIR managed through buffers and queue management for specific traffic.

\end{itemize}

By configuring the above parameters autonomously, ZTN can adjust the network parameters and perform closed-loop optimization in real time (Fig. \ref{congestion_control}). It monitors and measures the network state, analyzes the collected data, and determines actions to optimize bandwidth utilization. This process can be performed by the Q-learning agent adaptively. It can adjust the CA parameters to manage data flow rates based on UE traffic to maintain the desired QoS performance. Consequently, this should operate in a closed loop for perpetual optimization of network resources.

This paper proposes a novel closed-loop control mechanism for continuous optimization of network resources for ZTN.
Considering the network congestion scenario, with available bandwidth as state and traffic shaping as mitigating action, our proposed mechanism is able to predict the network state in terms of observed bandwidth with high accuracy as a first step. In the next step, based on the predicted network state an appropriate action in terms of traffic shaping parameters is taken autonomously so that the expected bandwidth matches with the actual value provided by the underlying network. This process forms the closed-loop control and optimizes network resources continuously. Simulation results in the section \ref{resul_discu} demonstrate that this closed-loop control can perform with a high accuracy of 95\%. To the best of our knowledge, the proposed scheme is probably one of the first implementation of a closed-loop control for ZTN.

The main contributions of this research work are as follows:
\begin{itemize}
  \item We propose a novel closed-loop control mechanism for the ZTN framework using congestion mitigation scenario.
  \item We develop a hybrid model which consists of Extreme Gradient Boosting (XGBoost) and Bidirectional Long Short Term Memory (BiLSTM) to accurately predict the network state that should match with the ground truth.
  \item We design a Q-learning algorithm to dynamically automate the decision-making process and take the appropriate action based on the predicted network state.

  \item We demonstrate the effectiveness of our proposed novel mechanism for closed-loop control in ZTN with extensive simulations.

\end{itemize}

The paper is organized as follows. Section II presents the system model. Section III discusses the results with simulation parameters. Section IV highlights the conclusions and future work.
\section{System Model}\label{section_system_model}
This section explains the system model which contains network topology, and closed-loop control using the XGBoosted BiLSTM model for network state prediction and Q-learning algorithm for action selection.
\subsection{Network Topology}

\begin{figure}[ht!]
    \centering
    \includegraphics[width=2.75in]{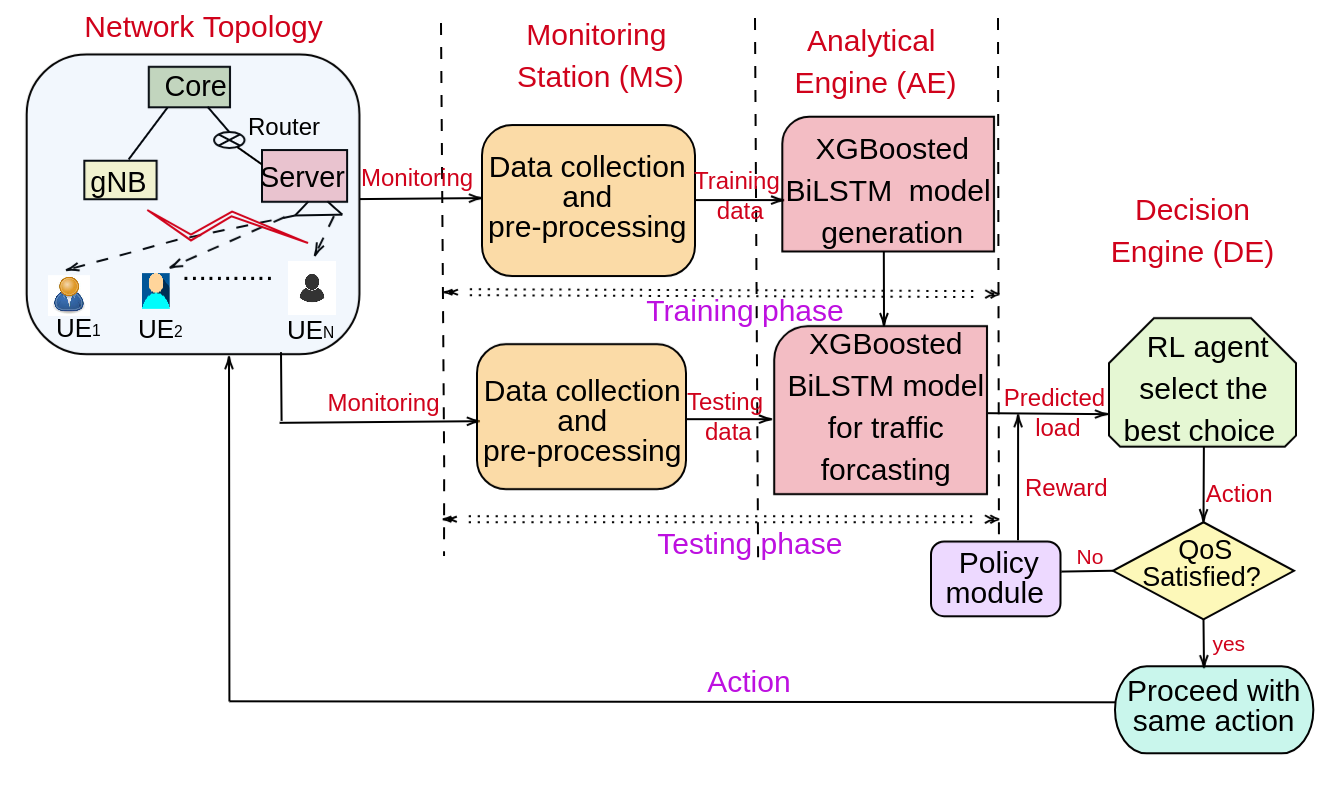}
    \caption{System model for closed-loop control mechanism}
    \label{System model}
\end{figure}

The network topology consists of a cellular network with the Core Network (CN) connected to $m$ application servers on the internet. $N$  UEs are served by $K$ base stations, i.e., Next Generation Nodes (gNBs), which are connected to CN networks as illustrated in Fig. \ref{System model}. Data transfers occur between the UEs and the application servers. Key QoS parameters of these data transfers, such as, \(bandwidth, latency, jitter,\) and \(packet loss\) can be measured and saved as a dataset.

\subsection{Network State Prediction}
The collected time-series dataset from the network topology helps to build the BiLSTM model \cite{cao2022} for the network state prediction. The BiLSTM layers have been trained in two directions using the dataset. In the forward path, training operations are performed on the original input samples. The backward path processes a reversed replica of the input data through BiLSTM layers. 
Then, the trained BiLSTM model takes the testing samples as input to predict the network state. Furthermore, the residuals ($\delta_{i}$) which represent the difference between the actual network state parameter and the respective BiLSTM predicted value are calculated using the formula given by equation (\ref{eqn_residue}).

\begin{equation}\label{eqn_residue}
    \delta_{i}= y_i - \hat{y}_i^{(BiLSTM)},       i= 1,2,.......,t
\end{equation}
here $y_i$, $\hat{y}_i^{(BiLSTM)}$ represent respectively, the actual parameter value, and BiLSTM predicted value for the $i$-th testing sample. For a testing set with $t$ samples, residuals vector ($\mathbf{res}$) is defined in equation (\ref{eqn_residue_vector}).
\begin{equation}\label{eqn_residue_vector}
\mathbf{res} = [\delta_1, \delta_2, \dots, \delta_t].
\end{equation}

To improve the prediction performance, the XGBoost model is trained with these residuals ($\mathbf{res}$). This allows the model to learn the BiLSTM prediction error pattern. Once trained, the XGBoost model produces its own set of predicted residuals, denoted as ${\hat{\delta}}^{i}$. It is used to adjust the initial BiLSTM prediction errors with equation (\ref{eqn_xgboost_adjust}).
\begin{equation}\label{eqn_xgboost_adjust}
    \hat{y}_i = \hat{y}_i^{(\text{BiLSTM})} + \hat{\delta}_i.
\end{equation}

The hybrid model (BiLSTM+XGBoost) predicts the QoS parameter value $\hat{y}_{i}$, much closer to the actual value $y_{i}$. The performance of the hybrid model is measured by the Mean Squared Error (MSE) metric given in equation (\ref{eqn_mse}).
\begin{equation}\label{eqn_mse}
  \text{MSE} = \frac{1}{t} \sum_{i=1}^{t} \left(y_i - \hat{y}_i\right)^2.
\end{equation}

\subsection{Optimal Action Selection}
The best action for the predicted network state is selected to adapt to dynamic network conditions, applying Q-learning which consists of the following attributes:
\begin{enumerate}
\item A state is defined as a tuple $\mathbf{s} = (s_1, s_2, ..., s_P)$ of $P$ QoS parameters in a state space ($\mathbf{s}$) representing the predicted network state by the BiLSTM model (output $\hat{y}_{i}$). For example, $s_1$ and $s_2$ can be the average bandwidth and jitter respectively, the network is currently experiencing based on the number of applications or UEs served by the network.

\item The ZTN management entities can program the network with different combinations of QoS parameters based on the reported underlying network state to optimize the same. Let, there be $Q$ such configuration parameters $c^{(n_1)}_1,
c^{(n_2)}_2, ..., c^{(n_Q)}_Q$. Each of these $c^{(n_i)}_i$ can in turn be sets of $n_i$ elements where $i = 1, 2,..,Q$. For example, $c^{(3)}_1$ can be the priority that can be assigned for the data traffic for a particular application. Since, $n_i = 3$ here, the set $c^{(3)}_1$ may be defined as  $c^{(3)}_1 = \{LOW, MEDIUM, HIGH\}$. Thus, a programmable configuration can be tuple $c \in$  $C = c^{(n_1)}_1 \times c^{(n_2)}_2 \times ... \times c^{(n_Q)}_Q$ and its corresponding cardinality is denoted with $|C|$.
\item The set of actions $A = \{a_1, a_2, ..., a_{|c|}\}$ is the choice of tuple $c$ from $C$, ZTN makes to optimize the network based on the underlying network state $s$ at time $t$. This is done by the Q-learning module which is explained in the section \ref{qlr_section}.
\end{enumerate}

\subsection{Illustration of a congestion mitigation action space}
Considering three CA parameters, the actions can be defined as follows.
For $c_1$, the bandwidth configuration can be modified based on the application services (Enhanced Mobile Broadband (eMBB), e.g., video, Ultra Reliable Low Latency Communications (URLLC), e.g., critical care and massive Machine-Type Communications (mMTC), e.g., sensors) as:
\begin{itemize}
\item Increase CIR for critical traffic during high congestion (reported by the network state).
\item Allow non-critical traffic in bursts using EIR when there is low congestion (reported by the network state).
\item Adjust buffer sizes to manage traffic bursts effectively with EBS.
\end{itemize}

For $c_2$ and $c_3$, different levels of priority and QoS model selection have been programmed based on the Third Generation Partnership Project (3GPP) specification \cite{3gpp2021} given in Table \ref{c2 and c3}. This table lists priority classifications and QoS models associated with different application types in mobile networks. Low-priority ($L$) applications may use a ``BE” model which ensures data is transmitted for small bursts (e.g. mMTC).
\begin{table}[ht!]
\centering
\caption{QoS Models for Traffic Management \cite{3gpp2021}}
\label{c2 and c3}
\begin{tabular}{|c|c|c|}
\hline
{\textbf{Application}} & {\textbf{Priority}} & {\textbf{QoS model}}              \\ \hline

FTP (mMTC)                            & $L$                                  & BE (Best Effort)           \\ \hline
Video (eMBB)                          & $M$                                  & RTPS (Real-Time Polling Service)        \\ \hline
Critical Care(URLLC)                         & $H$                                  & UGS (Unsolicited Grant Service) \\ \hline
\end{tabular}
\end{table}
Furthermore, video streaming under eMBB requires a Medium priority ($M$) level to achieve real-time responsiveness and utilize ``RTPS”. Specifically, URLLC may use a High-priority ($H$) and ``UGS” QoS model. The actions based on these CA configurations can manage resources to meet necessary QoS requirements during network congestion.

\subsection{Q-learning algorithm} \label{qlr_section}
Algorithm \ref{qlearning_algo} describes the closed-loop control mechanism for the ZTN implementation.

The expected performance metric ($E_Q$) can be kept fixed at a theoretical maximum value. For example, it can be the bandwidth when the channel condition is error-free. The aim of the closed loop ZTN framework is to move as close to $E_Q$ as possible to choose the best action for the predicted state.

The network state $O_Q$ is the output of the BiLSTM +XGboost model which is $\hat{y}_{i}$. The reward is defined as the negative squared difference between expected $E_Q$ and obtained metric $O_Q$ given in equation (\ref{reward_equ}). This squared difference approach magnifies the mismatches, leading to larger penalties for actions that deviate substantially from the expected performance.
    \begin{equation}\label{reward_equ}
         R = r(s, a) = -[(E_Q - O_Q)^2].
    \end{equation}

\begin{algorithm} [h!]
\caption{Q-Learning for QoS Optimization} \label{qlearning_algo}
\begin{algorithmic}[1]
\STATE \textbf{Input:} $states$ ($s$ $\epsilon$ $\mathbf{s}$), $actions$ ($a$ $\epsilon$ $\mathbf{A}$), $E_Q$, $O_Q$
\STATE \textbf{Initialize $q\_table$:} $Q(s,a) \leftarrow 0 $
\STATE \textbf{Define hyperparameters:}  $\alpha$, $\gamma$, $\epsilon_{\text{start}}$, $\epsilon_{\text{end}}$, $\epsilon_{\text{decay}}$, $\eta$

\FOR{$1$ to $\eta$}

    \STATE  $total\_reward \leftarrow 0$

    \FOR{$s$ in $\mathbf{s}$ }
        \STATE Choose an action ($a$) using $\epsilon$-greedy policy
        \STATE Get reward $r(s,a) \leftarrow$ equation \ref{reward_equ}.

        \STATE Update $q\_table  \leftarrow$ equation \ref{q-value}.

        \STATE Identify the best action using $q\_table$.
        \STATE Append the $total\_reward$ value.

    \ENDFOR
    \STATE Decay $\epsilon  \leftarrow$ equation \ref{epi_decay}.

\ENDFOR
\STATE \textbf{Output:} Best action that matches $O_Q$ with $E_Q$, MAE.
\end{algorithmic}
\end{algorithm}

The $q\_table$ matrix is initialized with zero. The column and row of this matrix are indexed by the action (a $\epsilon$ $\mathbf{A}$), and network state (s $\epsilon$ $\mathbf{s}$) respectively. Furthermore, hyper-parameters, such as learning rate ($\alpha$), discount factor ($\gamma$), and exploration-exploitation parameters ($\epsilon_{\text{start}}$, $\epsilon_{\text{end}}$, and $\epsilon_{\text{decay}}$) is assigned for information learning through exploiting actions.
For each episode ($\eta$), the algorithm estimates ``$R$” and Q-values are updated in $q\_table$ as follows:
    \begin{itemize}
        \item \textit{Action Selection}: An action is chosen from $\mathbf{A}$ for a network state $\mathbf{s}$ (using epsilon-greedy policy) that best matches $O_Q$ with $E_Q$.

        \item \textit{Reward Calculation}: The value $R$ for the selected $a_{|c|} \in A$ is calculated using equation (\ref{reward_equ}) and added to $total\_reward$ for each state in $\hat{y}_i$.
        \item \textit{Q-table Updation}: Then, the Q-value $Q(s, a)$ (where $s$ $\in$ $\mathbf{s}$, $a$ $\in$ $\mathbf{A}$) of each state and every possible action for $\hat{y}_i$ are updated in $q\_table$ by using the Bellman equation:
         \begin{align}\label{q-value}
         Q_N(s,a) \leftarrow Q_P(s,a) + \alpha \Big( r(s,a) \nonumber\\+ \gamma \max_{a'} Q(s', a') - Q_P(s,a) \Big).
        \end{align}
        In equation (\ref{q-value}), $Q_N$ represents the updated Q-value, while $Q_P$ is the current Q-value, and $\max_{a'} Q(s', a')$ is the maximum estimated future Q-value for the next state $s'$ and action $a'$. The parameter $\alpha$ controls the learning rate; $\alpha$=$0$ means No learning, a small $\alpha$ ensures slow learning, and a high $\alpha$ emphasizes new information. The $\gamma$ value balances the weight of immediate versus future rewards.

\item \textit{Best Action Selection}: The best action has been selected using $\epsilon$-greedy strategy,
\begin{equation}\label{epi_decay}
\epsilon_{\eta} = \epsilon_{\text{start}} \times (\epsilon_{\text{decay}})^{\eta},
\end{equation}
$\epsilon_{\text{start}}$ is the initial epsilon value. $\epsilon_{\text{decay}}$ is the decay rate, and $\eta$ represents the current episode. After each $\eta$, $\epsilon$ is gradually reduced until $\epsilon$ $>$ $\epsilon_{\text{end}}$ by multiplying it with $\epsilon_{\text{decay}}$ raised to the power $\eta$. This allows the Q-learning agent to shift progressively from exploration (random actions) to exploitation (choosing the best-known actions) as described in equation (\ref{epi_decay}) to match $O_Q$ close to $E_Q$.

    \end{itemize}

\begin{figure*}[ht!]
\centering
\subfigure[]{\includegraphics[width=2in]{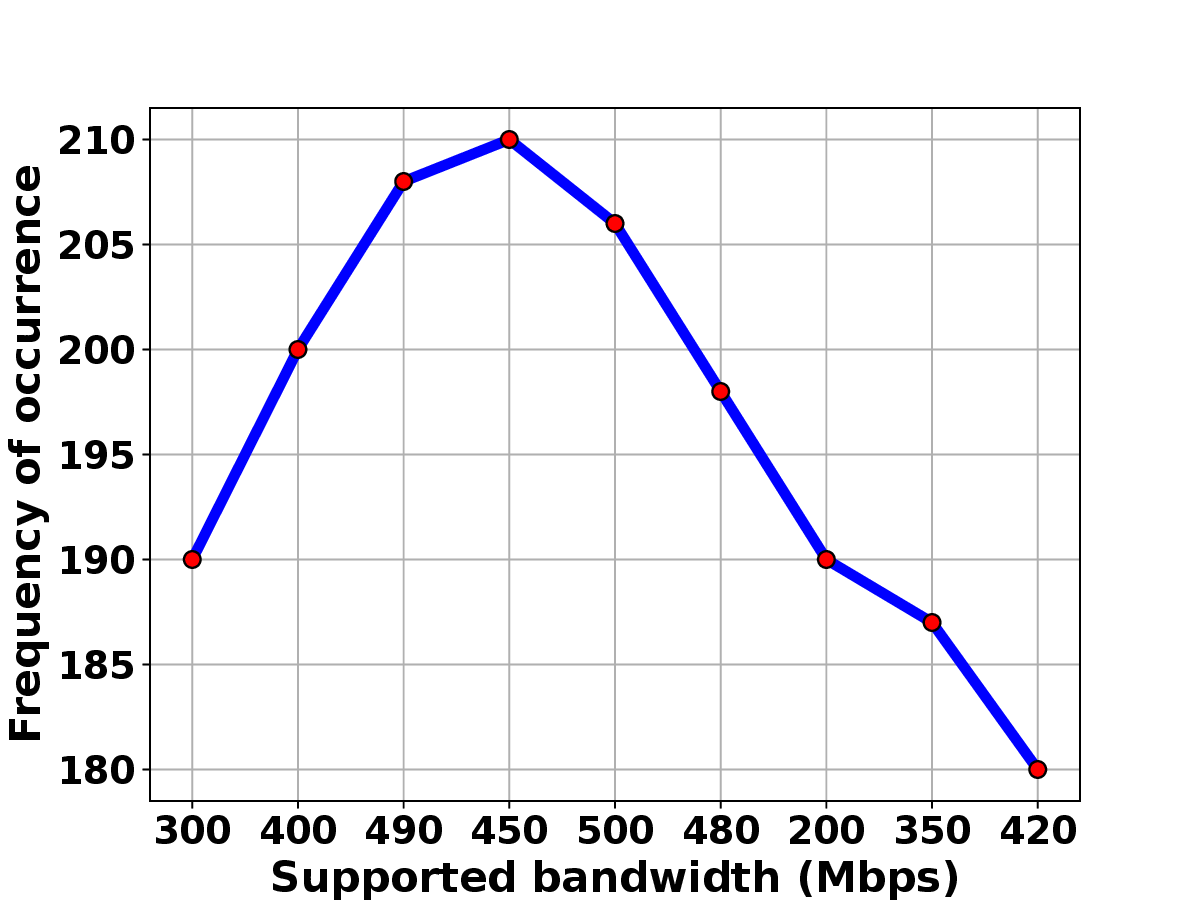}\label{traffic_dis}}
\hfil
\subfigure[]{\includegraphics[width=2in]{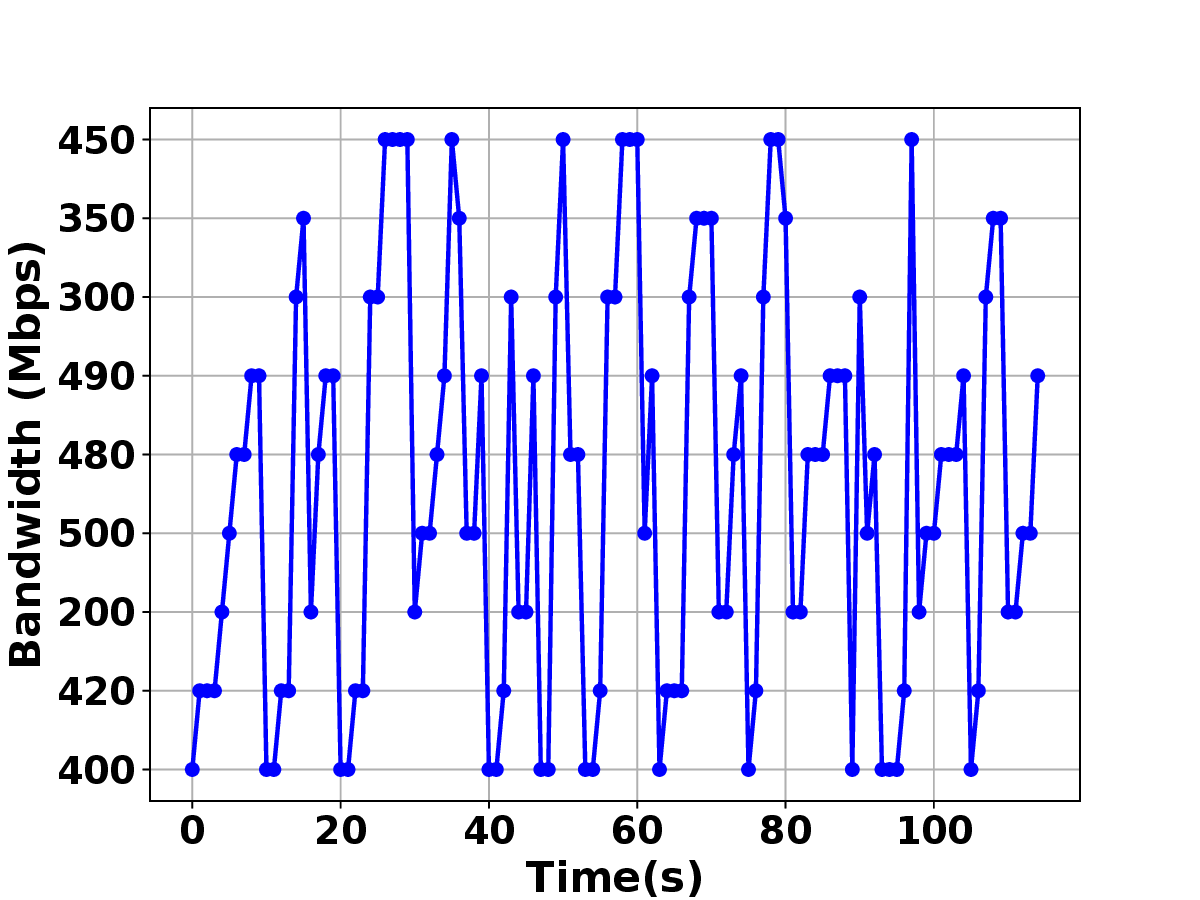}\label{traffic_pattern}}
\caption{Generated dataset for closed-loop control of ZTN framework (a) Traffic distribution and (b) Traffic pattern}
\end{figure*}

\section{Results and Discussion}\label{resul_discu}
This section presents the simulation results based on the system model and demonstrates the closed-loop ZTN function. To avoid complexity, this work only considers observed network Bandwidth (BW) as the QoS parameter, i.e., ${y}_i$ (which is also the same as $O_Q$ and the network state $\mathbf{s}$ as explained above). However, other parameters can be easily added to the performance evaluation of the proposed generalized system model.

\begin{figure*}[ht!]
\centering
\subfigure[]{\includegraphics[width=2in]{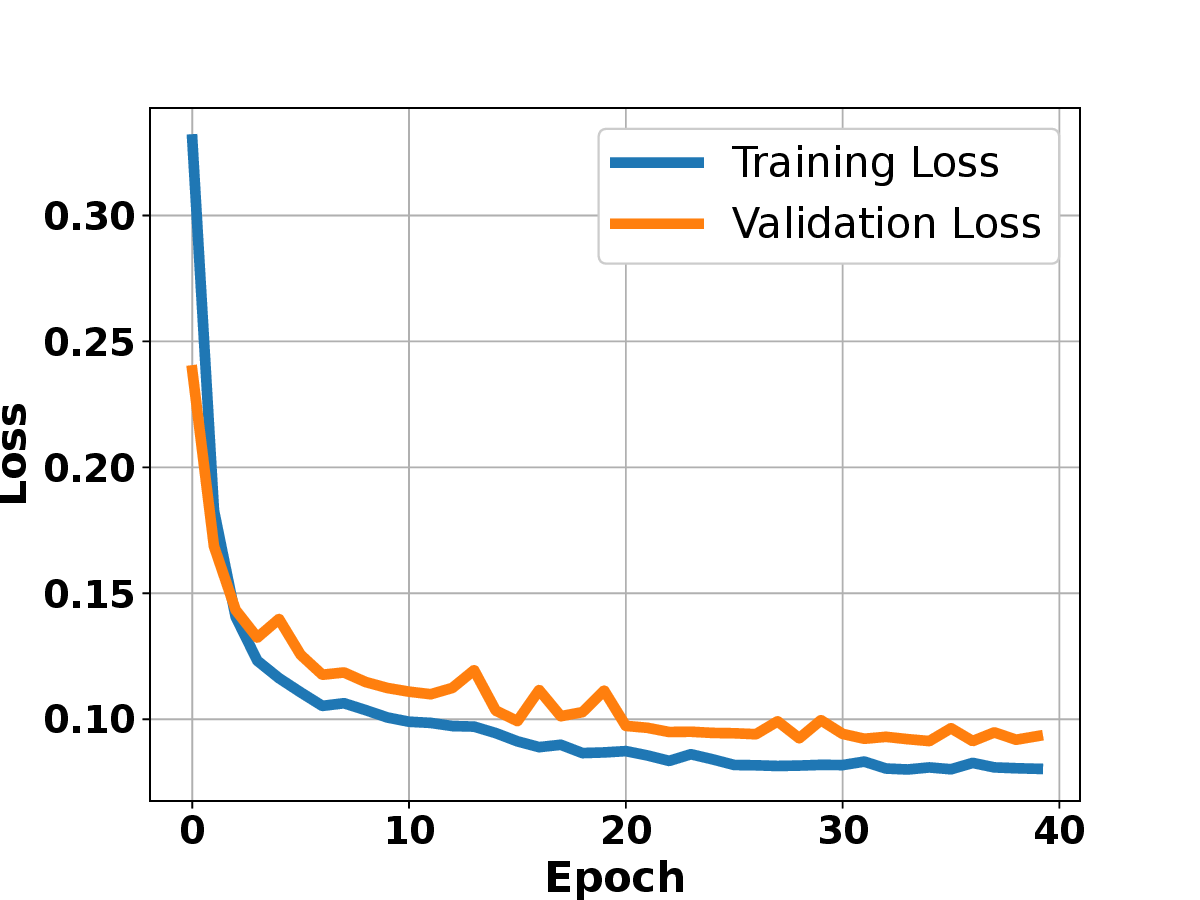}\label{lstm_model}}
\hfil
\subfigure[]{\includegraphics[width=2in]{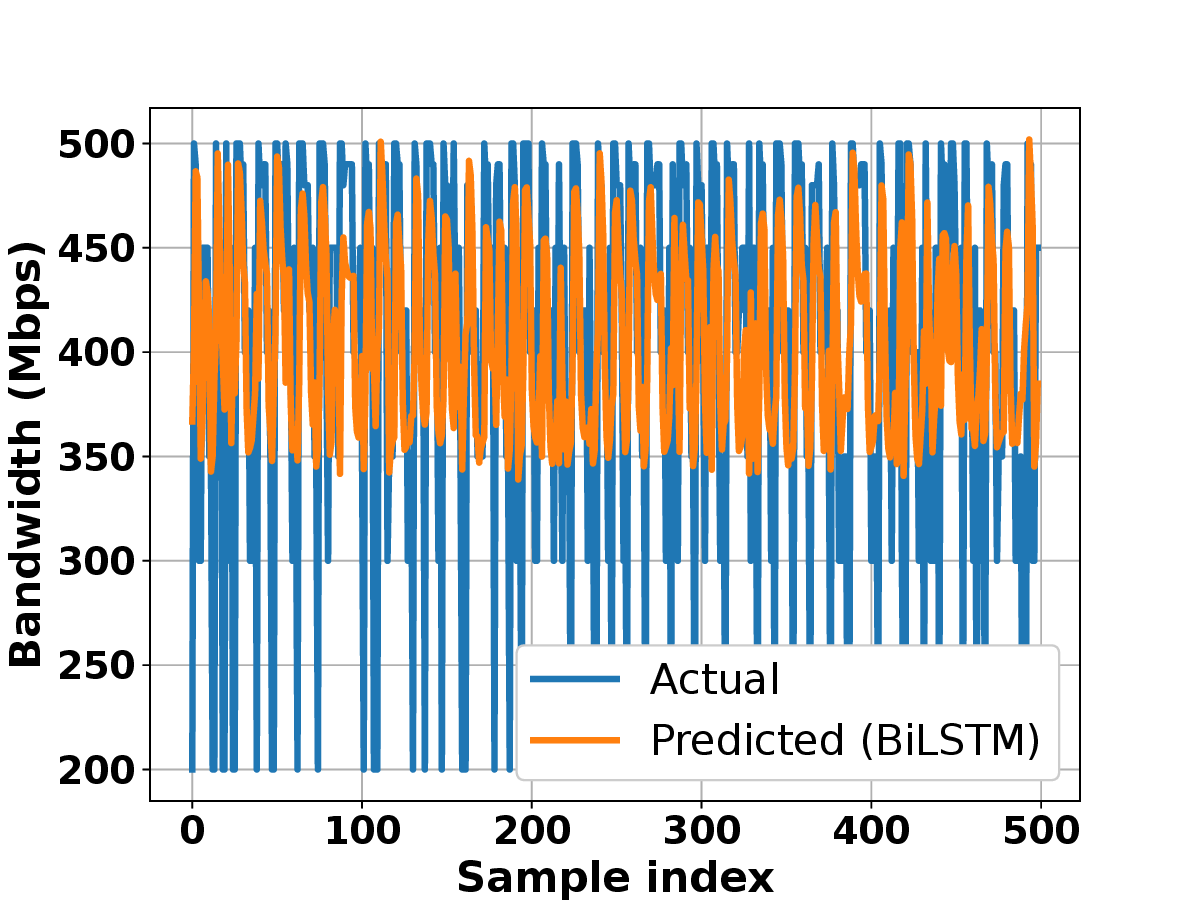}\label{lstm_model_pred}}
\hfil
\subfigure[]{\includegraphics[width=2in]{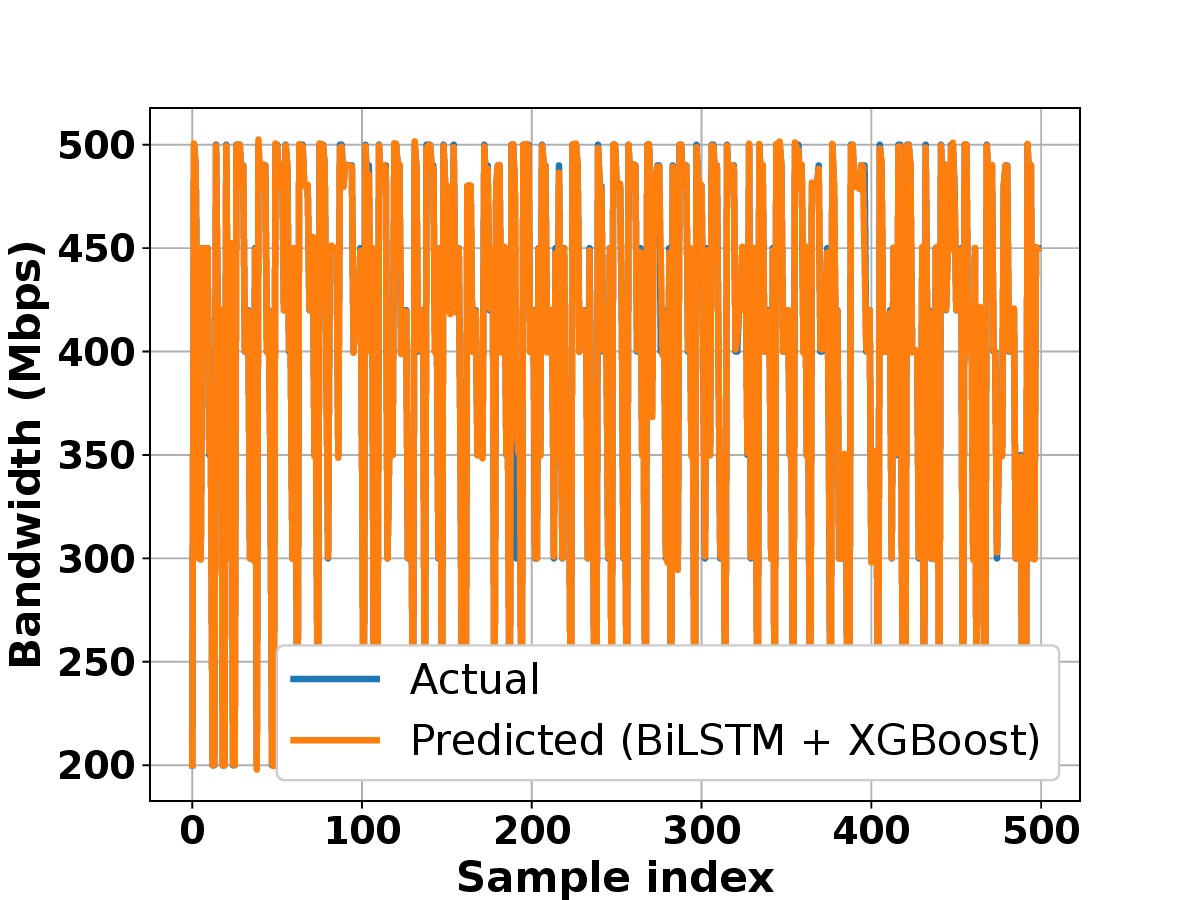}\label{hybrid_model_pred}}
\caption{Performance analysis of (a) BiLSTM training  (b)  BiLSTM prediction and (c) BiLSTM+XGBoost prediction}
\end{figure*}

\subsection{Experimental setup and dataset preparation}
The network topology for the E2E system is developed using the commercially available \textit{NetSim} \cite{tetcos2024} simulation tool. It consists of a CN that includes an Access and Mobility Management Function (AMF), Session Management Function (SMF), and User Plane Function (UPF) which connects to gNB. The RAN comprises a gNB and UEs with uplink and downlink data transfers to and from the Data Network (DN) through the UPF. The key parameters used in the simulation are summarized in Table \ref{simulation_parameters}. UE's frequency of occurrence (y-axis) into the network is random and follows a Poisson distribution with the mean occurrence rate ($\lambda$) against the observed bandwidth (along the x-axis) shown in Fig. \ref{traffic_dis}. As UEs arrive, the observed BW (y-axis) which supports different kinds of applications in the UEs is stored as a time series dataset depicted in Fig. \ref{traffic_pattern} over 100 seconds (x-axis). Since we are considering a network congestion scenario, it is injected into the network. The traffic shaping parameters are then configured to mitigate it. These observations are also saved and used for model training.

\begin{table}[ht!]
\centering
\scriptsize
\caption{Simulation parameters}
\label{simulation_parameters}
\begin{tabular}{|c|c|}
\hline
\textbf{Parameter}                                       & \textbf{Value} \\ \hline
Training and testing ambience   & i5 CPU,Tensorflow+Keras \\ \hline
\multicolumn{2}{|c|}{\textbf{NetSim platform}}                       \\ \hline
Number of gNB ($K$) & 1 \\ \hline
Total number of UEs ($N$) & 30 \\ \hline
Transmit power of gNB & 30 dBm \\ \hline
Channel bandwidth & 100 MHz \\ \hline
Numerology ($\mu$) & 1 \\ \hline

\multicolumn{2}{|c|}{\textbf{Hybrid(BiLSTM+XGBoost) model}}                       \\ \hline
Samples in training dataset    & 1800 \\ \hline
Samples in testing dataset  & 700  \\ \hline
Loss function & Mean square error \\ \hline
Optimizer & Adam \\ \hline
Activation function & ReLu \\ \hline
    $n\_{estimators}$ & 100 \\ \hline
    $learning\_rate\_{XG}$ & 0.5\\ \hline

\multicolumn{2}{|c|}{\textbf{Q-learning algorithm}}                       \\ \hline
Learning rate ($\alpha$) & 0.1 \\ \hline
$\epsilon_{\text{start}}$ & 1.0    \\ \hline
$\epsilon_{\text{decay}}$ & 0.995  \\ \hline
$\epsilon_{\text{end}}$   & 0.01   \\ \hline
Discount factor ($\gamma$) & 0.9 \\ \hline
No. of episodes ($\eta$)                                                  & 40,000 \\ \hline
\end{tabular}
\end{table}


\subsection{LSTM performance}
The BiLSTM model comprises three bidirectional LSTM layers with 50 cells each and dropout layers to prevent overfitting. Dense layers with ReLU activation predict future network BW. The BiLSTM model is trained using collected time series data samples, normalized using Min-Max Scaler. A step size 9 with a sliding-window approach has been applied to the training samples and reshaped into the BiLSTM input format. The model uses the Adam optimizer, trained and validated on 80\% and 10\% of the data samples, respectively. After 40 epochs, the fitness of the BiLSTM model is evaluated by plotting the training and validation losses (blue and orange lines, respectively) shown in Fig. \ref{lstm_model}. It is observed that the loss value (y-axis) decreases as the epochs (x-axis) increase. Additionally, it has been observed that the smooth line offers a good fit. The prediction performance of the trained BiLSTM model is tested using the remaining 10\% of the dataset depicted in Fig. \ref{lstm_model_pred}. It can be noticed that the MSE between the actual and predicted BW (y-axis), shown in the blue and orange lines respectively, is $40.3$, which is higher than expected indicating the need for further improvement. This MSE is considered as residuals ($\delta_{i}$) for the generation of the XGBoost model to improve the prediction performance (explained next).

\subsection{Hybrid (BiLSTM+XGBoost) model performance}
The XGBoost regressor model has been configured with the hyper-parameters from Table \ref{simulation_parameters}. It has been trained on $\delta_{i}$ to forecast $\hat{y}_i$. The performance of the hybrid model is shown in Fig. \ref{hybrid_model_pred}. It is evident that the predicted BW (orange line) is closely aligned with the actual BW (blue line). As a result, the MSE between the actual and predicted BW (y-axis) over a 500 sample index (x-axis) is $0.03$. Therefore, the XGBoost model significantly improves the prediction accuracy by compensating for the BiLSTM model errors.

\subsection{Q-learning for action selection}

\begin{figure}[ht!]
    \centering
    \includegraphics[width=2.4in]{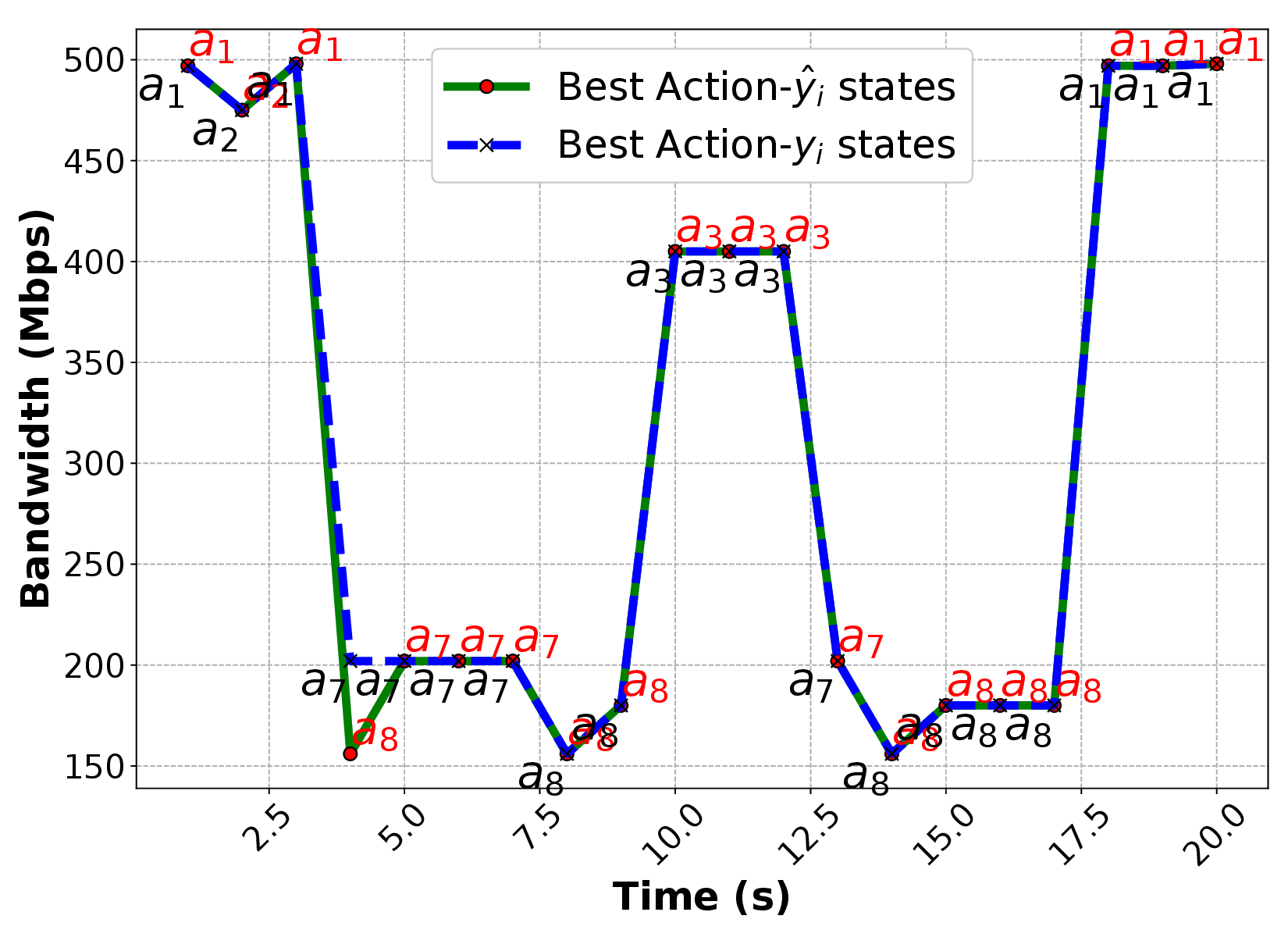}
    \caption{Effectiveness of Q-learning for selecting the best action}
    \label{RL_learning_pre}
\end{figure}

The Q-learning agent of ZTN recommends an action ($a_{|c|}$) from Table \ref{action_formulation} to optimize network performance under various congestion scenarios for the predicted BW state, $\hat{y}_i$. Each action corresponds to a specific combination of configuration parameters: $c^{(3)}_1$ (varied Generation Rate (GR)), $c^{(3)}_2$ (priority assignment: $L$, $M$, $H$), and $c^{(3)}_3$ (QoS model selection: BE, RTPS, UGS). The agent dynamically adjusts these parameters for three distinct applications: App1 (mMTC), App2 (eMBB), and App3 (URLLC) to select the action that optimizes the network performance. Fig. \ref{RL_learning_pre} illustrates the learning performance of the proposed Q-learning algorithm in selecting optimal actions. The green solid line with red circular markers represents the predicted network state, i.e., bandwidth, while the blue dashed line with black cross markers indicates the actual values. The annotations, such as $a_1$, $a_2$, $a_3$, $a_7$, and $a_8$, in black and red color, denote the specific action taken at various time instants (x-axis) for actual and predicted states (y-axis), respectively. The selected actions align closely for both $y_i$ and $\hat{y}_i$ states at all 20 timestamps except at time 4 seconds on the trained two-stage model which is 95\% accurate. Note that the actions based on the actual state are a reactive way to adjust the network performance whereas those for the predicted state are a proactive approach. The latter provides a faster response to changing network conditions.

\begin{table}[ht!]
\centering
\scriptsize
\caption{Action formulation using three CA parameters}
\label{action_formulation}
\begin{tabular}{|c|c|c|c|}
\hline
\textbf{Action} & \textbf{App1}     &\textbf{App2}    & \textbf{App3} \\ \hline
$a_1$  & GR        & $L$     & RTPS   \\ \hline
$a_2$  & GR        & RTPS    & $M$  \\ \hline
$a_3$  & $M$       & BE      & GR      \\ \hline
$a_4$  & $L$       & RTPS    & GR      \\ \hline
$a_5$  & RTPS      & $M$     & GR      \\ \hline
$a_6$  & $M$       & GR      & BE  \\ \hline
$a_7$  & $H$       & UGS     & GR      \\ \hline
$a_8$  & UGS       & GR      & $H$    \\ \hline
\end{tabular}
\end{table}

\begin{figure}[ht!]
    \centering
     \includegraphics[width=3.1in]{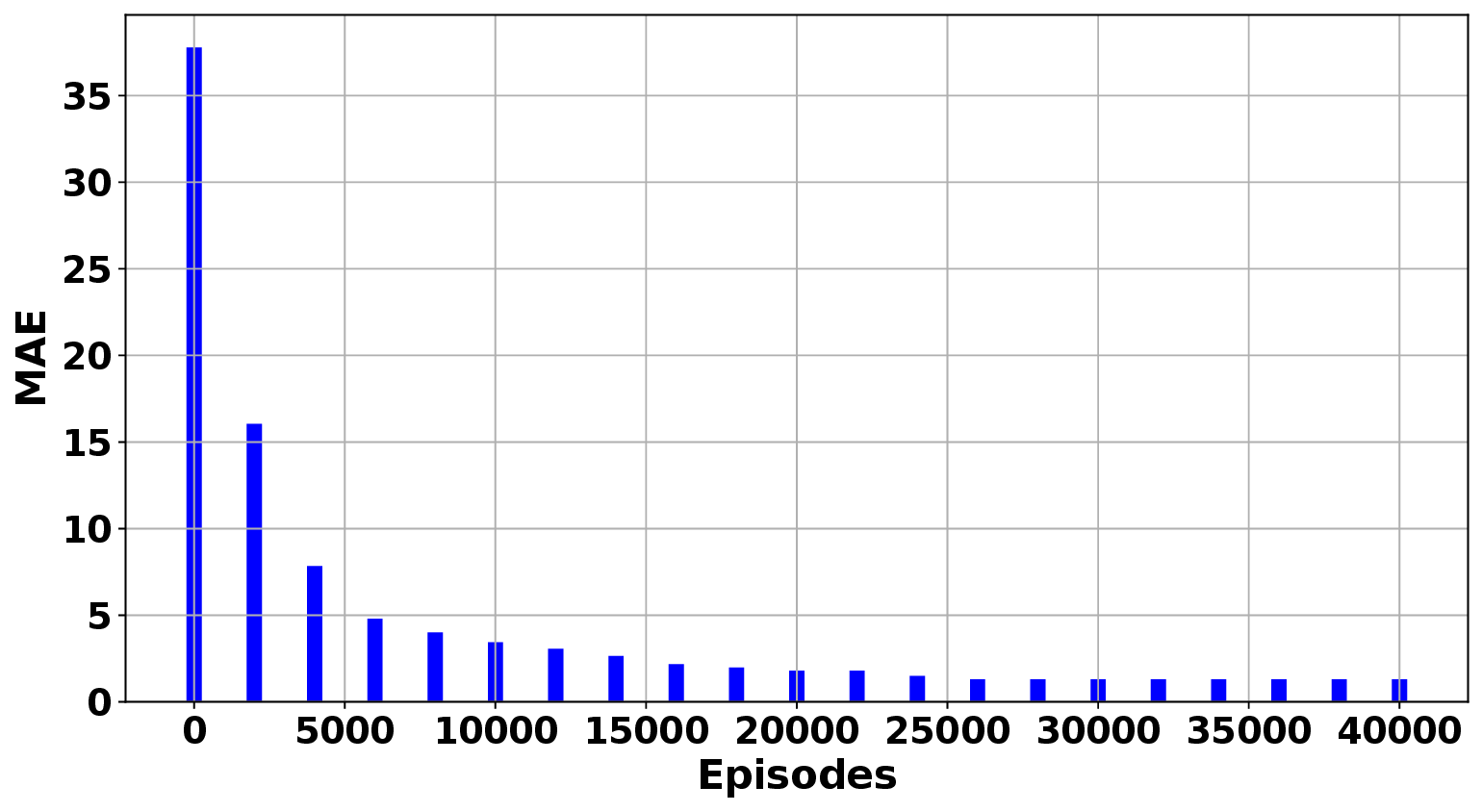}
    \caption{MAE performance of ZTN closed-loop control mechanism}
    \label{RL_conv_pre}
\end{figure}

The proposed Q-learning algorithm has been simulated over 40,000 episodes and compared for every 2,000 episodes shown in Fig. \ref{RL_conv_pre}. It has been observed that the MAE between the selected action of $y_i$ and $\hat{y}_i$ decreased to $\leq$ 1 from 37, as the number of episodes increased.

Thus, the results show that the expected bandwidth on choosing the optimal action for predicted state $\hat{y}_i$, matches the actual value $y_i$. This forms the closed-loop mechanism to optimize network performance perpetually and proactively under varying conditions.

\section{Conclusion}

Automation in 6G networks is critical for efficient functioning. We presented a novel closed-loop control for the ZTN framework to automate network management functionalities. Our approach is a two-stage mechanism. In the first stage, we employed a hybrid (BiLSTM+XGBoost) model to predict the network state. The second stage applied a Q-learning algorithm to select the optimal action for the predicted states. We considered the scenario of network congestion with available bandwidth as state and traffic shaping options as action to be taken. Simulation results have shown that the closed-loop control mechanism achieves 95\% accuracy in matching the actual network state by selecting the optimal action based on the predicted state. This work demonstrated a novel way to implement a ZTN closed-loop, a key topic largely unexplored. Future work will investigate extending this closed-loop mechanism for ZTN implementation for other scenarios, such as load balancing, etc.

\section*{Acknowledgment}
The authors would like to thank Toshiba Software India Pvt Ltd for sponsoring this research project.

\bibliographystyle{IEEEtran}
\bibliography{References.bib}

\end{document}